\documentclass[journal]{IEEEtran}
\usepackage{cite,amssymb}
\usepackage[pdftex]{graphicx}
\usepackage{algorithm}
\usepackage[cmex10]{amsmath}
\usepackage{textcomp}
\usepackage{algorithm}
\usepackage{algorithm}
\usepackage{algpseudocode}
\usepackage{array,balance,bm}
\usepackage{subfigure}

\ifCLASSOPTIONcompsoc
\else
\fi
\usepackage{url}
\usepackage{amsfonts}
\usepackage{exscale}
\usepackage{relsize}
\usepackage{multicol,color}

\newcommand{\mH}{\mathop{\rm H}}
\newcommand{\mT}{\mathop{\rm T}}
\begin{document}
	\title{Synthesis of Sparse Linear Arrays via Low-Rank Hankel Matrix Completion}
	\author{Xuejing Zhang,~\IEEEmembership{Member,~IEEE}
		%	\thanks{This work was supported in part by National Nature Science Foundation of China under Grants 61671139, 61771316 and 61701499, in part
		%		by Foundation of the Department of Education of Guangdong Province under Grant 2016KTSCX125, and in part by Nature Science Foundation of Guangdong Province 2017A030313341.}
		\thanks{X. Zhang is with the University of Electronic Science and Technology of China, Chengdu 611731, China (e-mail: xjzhang7@163.com).}}

\markboth{}%Ieee transactions on signal processing,~Vol.~13, No.~9, September~2016}%
{Shell \MakeLowercase{\textit{et al.}}: Bare Demo of IEEEtran.cls for Journals}
\maketitle

\begin{abstract}
	In the process of realizing the synthesis of the antenna array synthesis, it is of practical significance to arrange the nonuniform array and reduce the number of elements. According to the Matrix Pencil Method (MPM), we proposed an improved nonuniform array algorithm for reducing the number of array elements. We design the Hankel matrix by low-rank completion via log-det heuristic, and impose relatively loose constraints on the pattern, which may increase the degree of freedom of the radar array, so as to achieve the purpose of antenna array synthesis and non-uniform array. 
\end{abstract}


\IEEEpeerreviewmaketitle

\section{Introduction}
\IEEEPARstart{A}{rray} Antenna array pattern synthesis widely used in various fields, like radar, navigation, wireless communication and so on\cite{book}. Array synthesis involves many aspects of processing, reducing the number of array elements is one of the important aspects. In some specific applications, it is very important to synthesize the antenna array with the least number of elements to obtain the ideal beam pattern such as satellite communication. 

In this paper, we focus on reducing the element number of the linear antenna array. Many scholars have done related research on antenna array synthesis. Many scholars have done related research on antenna array synthesis. The research can be divided into two types: uniform array and non-uniform arraay. There are many traditional methods for uniform array synthesis such as Dolph-Chebshev and method\cite{2005}, which allows to synthesize narrow beam, low sidelobe or the optimization of an interesting parameter. Similar methods include Taylor method \cite{2004} and so on. The common defect of the traditional methods mentioned above is that there are too many elements needed, so the nonuniform array method emerges. The non-uniform array method can provide more degrees of freedom for the antenna array to generate the beam pattern we need, but also reduce the number of array elements and save radar resources \cite{1966}. This method can also be divided into several types. Firstly, some optimization algorithoms such as dynamic programming \cite{1964}, gentic algorithm (GA) \cite{1997}, particle swarm optimization (PSO) \cite{2007} and differential evolution algorithm (DEA) \cite{2003}. In reducing the number of array elements, these methods can find the global optimal solution, but it will take a lot of time. Secondly, some analytical methods \cite{1999,2005-2} and other synthesis algorithm \cite{1991,2007-2,2003-2,1988}. For these methods, they use a fixed number of elements to form an antenna array. They need to change the specified number of elements to find the one with less elements in all possible solutions. In the case of large number of elements, the calculation efficiency of this type of method will be low.

The matrix Rank Minimization Problem (RMP) is also a common problem and it is not easy to solve computationally \cite{1996}. RMP are encountered in many fields, such as signal processing, system identification , computational geometry and so on \cite{2002}. There are many heuristic algorithms to deal with this kind of problem, especially in the design of low-order controller \cite{1996-2,1994,1998}. 

Among the many methods to reduce the number of matrix elements through low-rank methods, there is a method called MPM \cite{2008-9}. This method first gives a reference pattern and samples it at equal intervals, and then achieves a non-uniform array by approximating the sampling points one by one and singular value decomposition (SVD), thereby reducing the number of elements. It is precisely because of its working principle that it also has certain defects. For example, there is no definite theoretical basis for the reference pattern. The need to approach all sampling points in the reference pattern will also reduce the degree of freedom of the array. We mainly improve the second defect of the MPM.

According to the idea of MPM, in this paper, we construct the ideal low rank Hankel matrix by using log det heuristic algorithm to complete the low rank, so as to achieve the purpose of non-uniform array and reduce the number of elements. Later, we will first describe the principle of our proposed algorithm, and then carry out simulation in different scenarios, and compare with the beam pattern and element distribution generated by MPM under the same conditions (the number of array elements is the same after non-uniform array).

\section{Problem Formulation}
%\subsection{Problem Formulation}
Considering a linear array with $ M $ isotropic elements, the beampattern radiated by the array can be expressed as
\begin{align}\label{key001}
F(u)={\bf w}^{\mT}{\bf a}(u)
\end{align}
where $ u={\rm sin}(\theta)\in[-1,1] $ with $ \theta $  representing the radiation angle, $ {\bf w}=[w_1,\cdots,w_M]^{\mT} $ is the vector of the excitation coefficients, $ {\bf a}(u)=[e^{j({2\pi}/{\lambda}){d_1}u},\cdots,e^{j({2\pi}/{\lambda}){d_M}u}]^{\mT} $ is the steering vector at $ u $, $j=\sqrt{-1}$, $\lambda$ is the wavelength, $ d_m $ is the position of the $ m $th element. 
It should be noted that the beampattern $ F(u) $ in \eqref{key001} is usually complex-valued.

To achieve a desirable radiation beampattern with reduced elements, the following (matching-based) sparse array synthesis criterion is formulated in as
\begin{subequations}\label{eq2n011}
	\begin{align}
	\label{eqn01202}
	\min_{\left\{\hat d_q,\hat w_q\right\}^Q_{q=1}}&~~Q\\
	\label{eqn03202}{\rm s.t.}&~\int_{-1}^{1}\Big| F_{\rm ref}(u)-\sum_{q=1}^{Q}\hat w_q e^{j({2\pi}/{\lambda})\hat d_qu}\Big|^2 du\leq\epsilon
	\end{align}
\end{subequations}
where $ F_{\rm ref}(u) $ is the preset reference pattern, $\epsilon$ is a small tolerance, ${\hat d_q}$ and ${\hat w_q}$ are the positons and excitations to be optimized, $ q=1,\cdots,Q $. 

In the above problem \eqref{eq2n011}, an appropriate reference pattern is needed for a satisfactory synthesis and different choices of reference patterns may lead to different synthesis results. Moreover, one can see from 
\eqref{eqn03202} that both the amplitude responses and phase responses are considered in the constraint of pattern matching, and the matching is conducted on both mainlobe and sidelobe regions. 
Since the phase distributions have no impact on the radiation power, the criterion in \eqref{eq2n011} occupies redundant design freedom and has limited capability on accurate sidelobe control.

In this paper, we propose a new criterion in sparse array synthesis by modifying \eqref{eq2n011} as 
\begin{subequations}\label{eq2nd011}
	\begin{align}
	\label{eqn01d202}
	\min_{\left\{\hat d_q,\hat w_q\right\}^Q_{q=1}}&~~Q\\
	\label{eqn0d3202}{\rm s.t.}
	&~\int_{\Omega_M}\Big| F_{\rm ref}(u)-\sum_{q=1}^{Q}\hat w_q e^{j({2\pi}/{\lambda})\hat d_qu}\Big|^2 du\leq\epsilon\\
	\label{eqnd0d3202}&~\Big|\sum_{q=1}^{Q}\hat w_q e^{j({2\pi}/{\lambda})\hat d_qu}\Big|^2\leq {\rho}(u),~u\in\Omega_S
	\end{align}
\end{subequations}
where $ {\Omega_M} $ and $ {\Omega_S} $ represent the mainlobe and sidelobe regions of the radiation pattern, respectively. Different from \eqref{eq2n011}, only the beampattern of mainlobe region is matched to a reference pattern $ F_{\rm ref}(u) $ as formulated in \eqref{eqn0d3202}.
For the sidelobe region, we constrain its power level to be lower than the upper bound $ \rho(u) $, thus leaving more freedom
for sparse array synthesis comparing with the matching constraint in \eqref{eqn03202}.

Note that the problem \eqref{eq2nd011} is intractable due to its sparsity on cost function. To solve problem \eqref{eq2nd011} and achieve sparse array synthesis, a low-rank hankel matrix completion method is proposed, as 
explained more detailedly next.

\section{The Proposed Algorithm}

Low-Rank Hankel Matrix Completion

Then, sample the pattern function with equal interval from $u=-1$ to $1$, the sampling interval (sampling u) is $\Delta$. Record the obtained sampling point (2N+1 points totally) as $x(n)$
\begin{align}\label{eq1}
x(n)&=F(u)|_{u=n\Delta}
\end{align}
where $n=-N,-N+1,\cdots,N$.

We firstly define a hankelization formula as
\begin{align}
&\mathcal H_{(S)}\left\{a_0,a_1,\cdots,a_i\right\}\nonumber\\
&=\left[
\begin{matrix}
a_0&a_1&\cdots&a_S\\
a_1&a_2&\cdots&a_{S+1}\\
\vdots&\vdots&\ddots&\vdots\\
a_{i-S}&a_{i-S+1}&\cdots&a_i
\end{matrix}
\right]\in{\mathbb C}^{(i-S+1)\times (S+1)}
\end{align}

\begin{figure}[!t]
	\centering
	\includegraphics[width=3.5in]{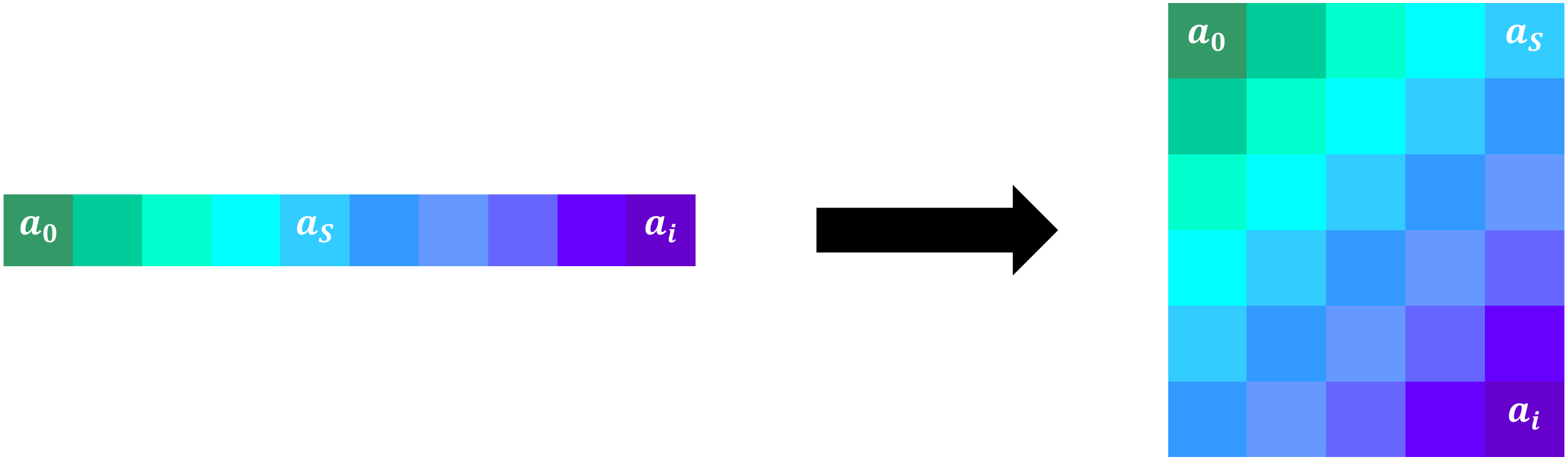}
	\caption{An abbraviated illustration of hankelization.}
	\label{fig1}
\end{figure}

For the convenience of understanding, we give a relatively intuitive diagram Fig.\ref{fig1}, where we can clearly see how $\mathcal{H}$ transformed the aggregate $\left\{a_0,a_1,\cdots,a_i\right\}$ into Hankel matrix with the parameter $S$.

The Hankel matrix ${\bf Y}$ can be constructed by the sampling points obtained above which can be denoted as 
\begin{align}\label{eq7}
{\bf Y}&={\mathcal H}_{(L)}\left\{x(-N),x(-N+1),\cdots,x(N)\right\}\nonumber\\
&\in{\mathbb C}^{(2N-L+1)\times (L+1)}
\end{align}
where $L$ is the matrix pencil parameter whose range is defined as $2N-L\geq M$ and $L+1\geq M$. Besides, we have 
\begin{equation}
{\rm rank}({\bf Y})=M
\end{equation}

Therefore, the problem of reducing the number of elements is equivalent to reducing the rank of Hankel matrix $\bf Y$.

After expounding the principle of MPM, it is not difficult to find that if we want to continue to generate the ideal beam, we need a reference pattern and approach each sampling point on the reference pattern. At the same time, it also shows the disadvantage of MPM, which is the dependence on reference pattern. In this section, we will propose an algorithm to design ideal low rank Hankel matrix with only a few reference sampling points. 

\subsection{The proposed algorithm model}
According to the idea of MPM, the problem of reducing the number of array elements can be transformed into RMP under specified constraints. It is known that each sampling data corresponds to an anti-diagonal in ${\bf Y}$. However, it is not necessary to approxiamte every point to the sampling point if the reference pattern which will waste freedom of the array. We can just approach sampling points on mainlobe. In other words, we construct the Hankel matrix ${\bf Y}$(the new hankel matrix is denoted as ${\bf Y}_R$) by approaching its several certain anti-diagonals to aprroach via log-det heuristic instead of every anti-diagonals.
In this paper, we might set a restraint response level $\rho$ (not necessary a constant value) to constrain the sidelobe of the reference beam by adding a Chebshev window as well. For the sampling points on the mainlobe, what we need to approach are the points over $\rho$. In order to obtain the desired sampling points, we start with the central sampling point of the mainlobe and the leftmost and rightmost sampling points of the main lobe are represented as $n_l$ and $n_r$. So that the mainlobe region $\mathbb{V}$ and sidelobe region $\mathbb{\overline V}$ can be denoted as
\begin{align}
\mathbb{V}&=\left\{n_l,\cdots,n_r\right\}\\
\mathbb{\overline V}&=\left\{-N,\cdots,n_l-1\right\}\cup\left\{n_r+1,\cdots,N\right\}
\end{align} 

After that, we approach these desired sampling points and limit the level of the sidelobe when constructing ${\bf Y}_R$ by designing the value of every anti-diagonal of ${\bf Y}_R$. 

Then the problem of reconstructing the hankel matrix ${\bf Y}$ can be turned into a RMP which can be shown as
\begin{align}\label{eq3}
min~~&{\rm rank}~{\bf Y}\nonumber\\
s.t.~~~&\left\{
\begin{aligned}
&{\bf Y}={\mathcal H}_{(L)}\left\{x(-N),x(-N+1),\cdots,x(N)\right\}\\
&\lvert x_R(n)-x(n)\rvert\leq\epsilon,~~n\in\mathbb{V}\\
&\lvert x_R(n)\rvert\leq \rho,~~n\in\mathbb{\overline V}
\end{aligned}
\right.
\end{align}
where $n=0,1,\cdots,N-1$, $\theta_0$ denotes the direction of the mainlobe, $x_R(n)$ stands for the value in the $n$th anti-diagonal of ${\bf Y}_R$, and $\epsilon$ denotes a small positive number.
After establishing the algorithm model, in the next subsection, we will cover the rank minimization issues of $\bf Y$.

\subsection{Rank minimization}
According to the semidefinite embedding lemma, the optimization problem in \eqref{eq3} can be rewritten as
\begin{align}\label{eq4}
min~~&\frac{1}{2}{\rm rank}~{\rm diag}({\bf P},{\bf Q})\nonumber\\
s.t.~~~&\left[
\begin{matrix}		
{\bf P}&{\bf Y}\\
{\bf Y}^{\mT}&{\bf Q}
\end{matrix}\right]\geq 0\nonumber\\
&\left\{
\begin{aligned}
&{\bf Y}={\mathcal H}_{(L)}\left\{x(-N),x(-N+1),\cdots,x(N)\right\}\\
&\lvert x_R(n)-x(n)\rvert\leq\epsilon,~~n\in\mathbb{V}\\
&\lvert x_R(n)\rvert\leq \rho,~~n\in\mathbb{\overline V}
\end{aligned}
\right.
\end{align}
where ${\bf P}={\bf P}^{\mT},{\bf Q}={\bf Q}^{\mT}$.

If we use the function ${\rm log~det}({\bf Y}+\delta{\bf I})$ as a smooth surrogate for ${\rm rank}~{\bf Y}$ in \eqref{eq3} which is called log-det heuristic, the objective function in \eqref{eq4} can be written as
\begin{equation}\label{eq6}
{\rm log~det}({\rm diag}({\bf P},{\bf Q})+\delta{\bf I})
\end{equation}

Consider about the function ${\rm log~det}({\bf Y}+\delta{\bf I})$, it is not convex. Therefore, we use the iterative linearization to deal with it. Hence, we utlize the first-order Taylor series expansion to express it as
\begin{equation}\label{eq5}
{\rm log~det}({\bf Y}+\delta{\bf I})\approx{\rm log~det}({\bf Y}_k+\delta{\bf I})+{\bf Tr}({\bf Y}_k+\delta{\bf I})^{-1}({\bf Y}-{\bf Y}_k)
\end{equation}
where ${\bf Y}_k$ denote the $k$th iterate of ${\bf Y}$, $k=0,1,\cdots,K$.

It is noticed that, we can treat the problem as a convex one when ${\bf Y}>0$. Then we can also get the new optimal ${\bf Y}_{k+1}$ as
\begin{equation}
{\bf Y}_{k+1}={\rm argmin}~{\bf Tr}({\bf Y}_k+\delta{\bf I})^{-1}{\bf Y}
\end{equation}
where we ignore some constant in \eqref{eq5} which will not affect the result.

On the basis of establishing the algorithm model, according to the semidefinite embedding lemma and log-det heuristic in \cite{2003-6}, the optimization problem in \eqref{eq3} can be rewritten as
\begin{align}\label{mod}
&{\rm diag}({\bf P}_{k+1},{\bf Q}_{k+1})=\nonumber\\
argmin~~&{\rm Tr}[({\rm diag}({\bf P}_k,{\bf Q}_k)+\delta{\bf I}^{-1}){\rm diag}({\bf P},{\bf Q})]\nonumber\\
s.t.~~~&\left[
\begin{matrix}		
{\bf P}&{\bf Y}\\
{\bf Y}^{\mT}&{\bf Q}
\end{matrix}\right]\geq 0\nonumber\\
&\left\{
\begin{aligned}
&{\bf Y}={\mathcal H}_{(L)}\left\{x(-N),x(-N+1),\cdots,x(N)\right\}\\
&\lvert x_R(n)-x(n)\rvert\leq\epsilon,~~n\in\mathbb{V}\\
&\lvert x_R(n)\rvert\leq \rho,~~n\in\mathbb{\overline V}
\end{aligned}
\right.
\end{align}
where ${\bf P}_0={\bf I}^{(2N-L+1)\times(2N-L+1)}$, ${\bf Q}_0={\bf I}^{(L+1)\times(L+1)}$ and $k=0,1,\cdots,K$.

According to this principle, we are able to reconstruct the new hankel matrix ${\bf Y}_R$ to replace the ${\bf Y}$ in \eqref{eq7} through $t$ iterations which is set to $K=10$ in the following simulations. With the well designed low rank Hankel matrix ${\bf Y}_R$, we can estimate the positions of the new elements and corresponding weights in the next subsection.

\subsection{Estimate the new element positions}
After constructing the new hankel matrix  ${\bf Y}_R$, with ${\rm rank}({\bf Y}_R)=R$, we can use the MPM to estimate the new antenna array. In the subsection above, we all know that the matrix pencil parameter $L$ needs to satisfy $M\leq L\leq 2N-M+1$. In this paper, we might set $2M=L=(N-1)/2$ as well. 

We can obtain two new matrix ${\bf Y}_{R1}$ and ${\bf Y}_{R2}$ by removing the first column and the last column respectively which are denoted as

\begin{align}
{\bf Y}_{R1}&={\mathcal H}_{(L-1)}\left\{x(-N+1),x(-N+2),\cdots,x(N)\right\}\\
{\bf Y}_{R2}&={\mathcal H}_{(L-1)}\left\{x(-N),x(-N+1),\cdots,x(N-1)\right\}
\end{align}
Then we can obtain the new element distributin by doing eigenvalue decomposition to ${\bf Y}_{R1}^\dagger{\bf Y}_{R2}$. The position of the $r$th element can be expressed as 
\begin{equation}
\hat{d_r}=\frac{\lambda {\rm ln}(\hat z_r)}{j2\pi\Delta}
\end{equation}
where $r=1,2,\cdots,R$, $\hat z_r$ denotes the eigenvalue of ${\bf Y}_{R1}^\dagger{\bf Y}_{R2}$ \cite{1995-2}. The relationship between $\hat z_r$, $\hat\omega_r$ and $x(n)$ is shown as 	
\begin{equation}
\underbrace{
	\left[
	\begin{matrix}
	\hat z_1^{-N}&\hat z_2^{-N}&\cdots&\hat z_R^{-N}\\
	\hat z_1^{-N+1}&\hat z_2^{-N+1}&\cdots&\hat z_R^{-N+1}\\
	\vdots&\vdots&\ddots&\vdots\\
	\hat z_1^{N}&\hat z_2^{N}&\cdots&\hat z_R^{N}
	\end{matrix}
	\right]}_{\hat{\bf Z}}
\underbrace{
	\left[
	\begin{matrix}
	\hat\omega_1\\\hat\omega_2\\\vdots\\\hat\omega_R
	\end{matrix}
	\right]}_{\hat{\bf w}_R}
=\underbrace{
	\left[
	\begin{matrix}
	x(-N)\\x(-N+1)\\\vdots\\x(N)
	\end{matrix}
	\right]}_{{{\bf x}}}
\end{equation}
Then we can solve the weight vector $\hat{\bf w}_R$ by the LS method as
\begin{equation}
\hat{{\bf w}}_R=(\hat{\bf Z}^{\mH}\hat{\bf Z})^{-1}\hat{\bf Z}^{\mH}{\bf x}
\end{equation}
Hence, till now, we obtain the new beampattern as 
\begin{equation}
\hat{f}(\theta)=\hat{{\bf w}}_R^{\mH}{\bf a}(\theta)
\end{equation}
The general algorithm flow is given in Algorithm \ref{al1}.

\begin{algorithm}[!t]
	\caption{Algorithm for synthesis of sparse linear array via low-rank hankel matrix completion}\label{al1}
	\begin{algorithmic}[1]
		\State {\bf Input:} $M,N,L,\rho,K,\epsilon$
		\State $f(\theta)\triangleq F(u)$
		\State $x(n)=F(u)|_{u=n\Delta}$, where $n=-N,-N+1,\cdots,N$
		\For {$k = 0,1,\cdots,K$} \eqref{mod}
		\EndFor	
		\State ${\bf Y}_R$
		\State ${\bf Y}_{R1}$, ${\bf Y}_{R2}$
		\State ${\bf Y}_{R1}^\dagger{\bf Y}_{R2}=\hat{\bf U}\Lambda\hat{\bf U}^{-1}$, where
		$\Lambda={\rm diag}\left\{\hat z_1, \hat z_2,\cdots, \hat z_R\right\}$
		\State $\hat{d_r}=\frac{\lambda {\rm ln}(\hat z_r)}{j2\pi\Delta}$, where $r=1,2,\cdots,R$
		\State $\hat{{\bf w}}_R=(\hat{\bf Z}^{\mH}\hat{\bf Z})^{-1}\hat{\bf Z}^{\mH}{\bf x}$
		\State $\hat{f}(\theta)=\hat{{\bf w}}_R^{\mH}{\bf a}(\theta)$
	\end{algorithmic}
\end{algorithm}

\section{Simulation}
In the sections above, we have already demonstrated our proposed algorithm and its difference with MPM. In this section, we will simulate the MPM method and our algorithm under two different side lobe constraints, and compare the pattern and element position generated by the two methods, as well as the simulation under changing the initial element conditions and side lobe constraint level.
\subsection{Conventional Chebshev weighted}
Consider a uniform linear array with $M=20$ elements as the reference array, with the mainlobe direct to 0, the element spacing is $\frac{\lambda}{2}$ and a -30dB Chebshev window. In order to enhance the performance of the algorithm, try to ensure that the sidelobe level is strictly lower than the sidelobe constraint, we can increase the number of samples appropriately. In this simulation, the sampling number is $2N+1=81$ which may ensure the good effect of the algorithm, the matrix pencil parameter is $L=40$ which may make sure ${\bf Y}_{\bf R}$ is a square matrix so as to ensure the stability of the algorithm and the number of iterations is set to $K=10$. 

For convenience of calculation and the stability of the algorithm, we construct the hankel matrix ${\bf Y}_R$ as a square matrix.

\begin{figure}[!tpb]
	\centering
	\subfigure[]
	{\includegraphics[width=3.5in]{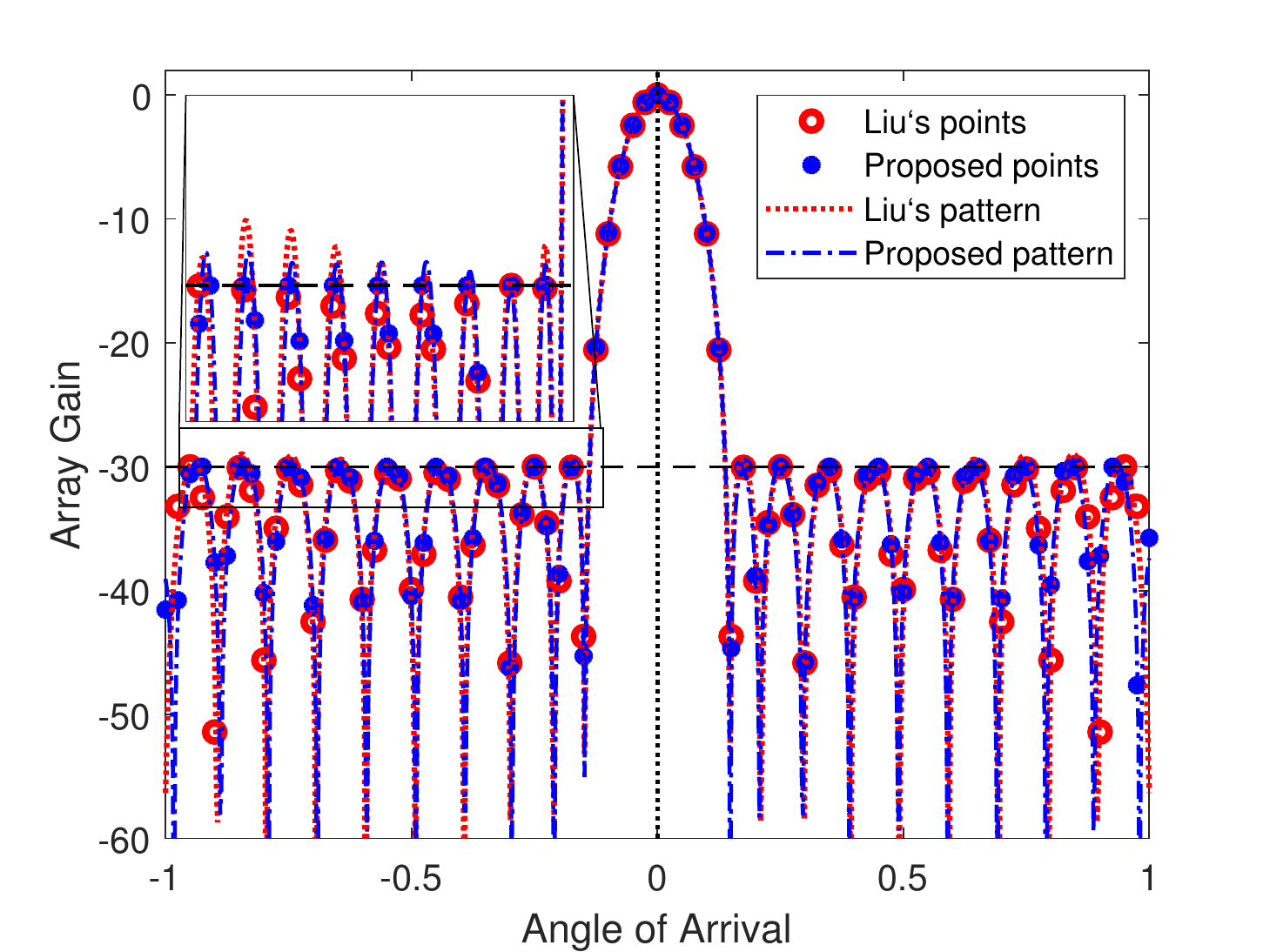}%
		\label{beamcom}}
	\ \
	\subfigure[]
	{\includegraphics[width=3.5in]{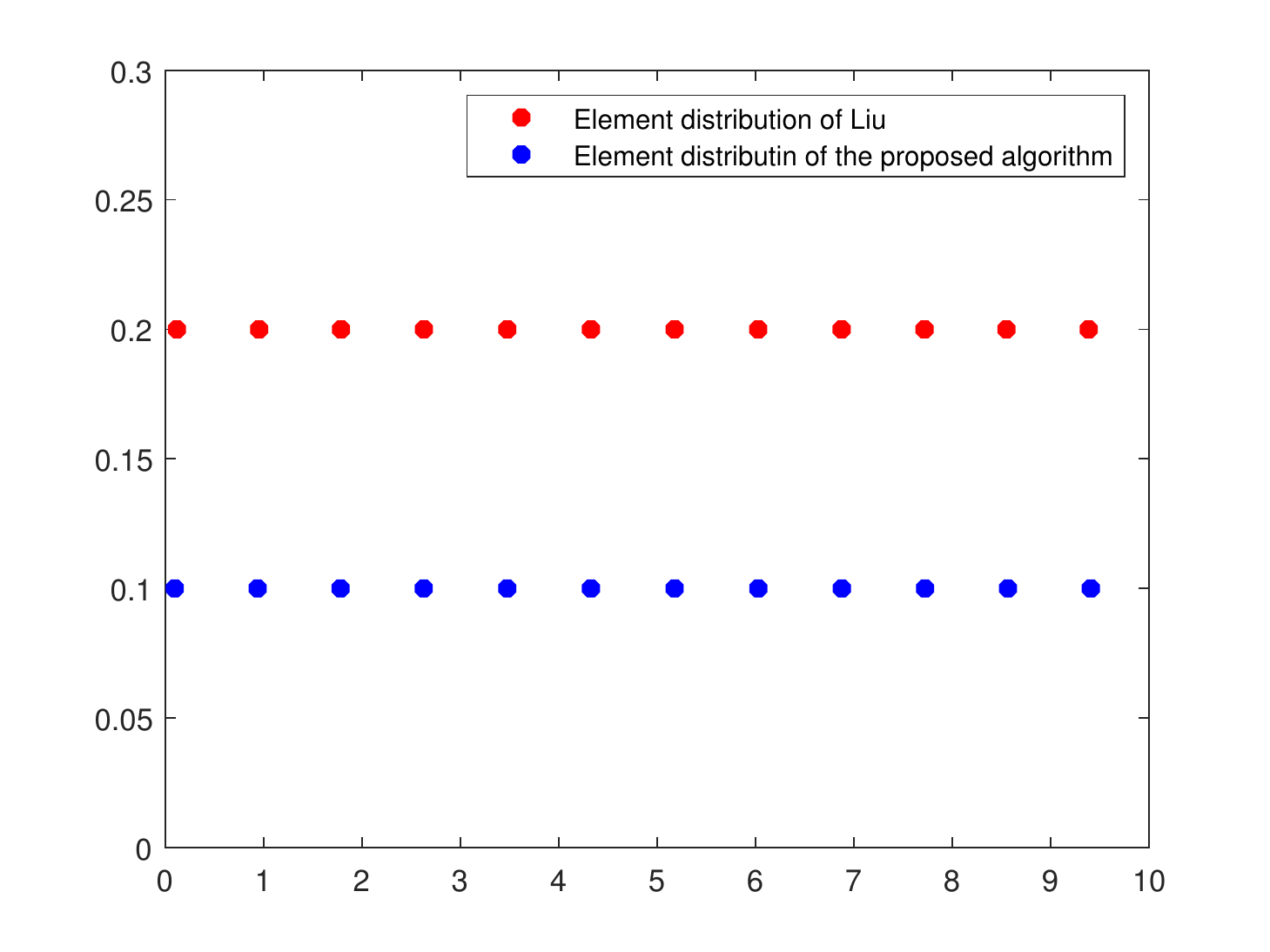}%
		\label{elecom}}			
	\caption{Comparision between Liu's method and the proposed method.
		(a) Beam pattrens.
		(b) Elements distribution (both are 12 elements).}
	\label{fig2}
\end{figure}

Besides, the fewer points we approach, the higher the degree of freedom of the array.
In this situation, we select the 41th sampling point as the central points, and the 40th, the 43th sampling point to be approached which can consider the performance of the left and right sides of the main lobe at the same time. The other points on the main lobe are not constrained, then the comparision with Liu's method is given in Fig.\ref{fig2}

In Fig.\ref{fig2}, both of the algorithms can generate ideal beampattern, meanwhile, both methods can reduce the number of array elements from 20 to 12. Fig.\ref{beamcom} shows the pattern generated by these two methods, it can be seen that the mainlobes of the two patterns are almost coincident, however, due to the relatively loose constraints, the proposed method has a higher freedom, and the sidelobe constraint effect of the generated pattern is better than that of the MPM. 

In order to show the array structure rearranged after calculation more clearly, we list the positions of the 12 new elements generated by two methods, where $n$ denotes the serial number of elements, $\hat{d_n}$ denotes the position of elements generated by our proposed algorithm, $d_n$ denotes denotes the position of elements generated by MPM. Combine Fig.\ref{elecom} and TABLE \ref{table1}, we can clearly see the position of each element and their comparision.

\begin{table}[!t]
	\renewcommand{\arraystretch}{1.5}
	\caption{The table of element position and weight}
	\label{table1}
	\centering
	\begin{tabular}{c | c | c || c | c | c}
		\hline
		$ n $ & $ \hat{d_n} $ & $ d_n $  
		&$ n $ & $ \hat{d_n} $ & $ d_n $\\
		\hline	
		1 & 0.0944  &0.1160 &7&5.1755&5.1753\\	
		2 & 0.9366  &0.9519 &8&6.0261&6.0252\\
		3 & 1.7796	&1.7846 &9&6.8749&6.8728\\
		4 & 2.6251  &2.6272 &10&7.7204&7.7154\\
		5 & 3.4379  &3.4748 &11&8.5634&8.5481\\
		6 & 4.3245  &4.3247 &12&9.4056&9.3840\\
		\hline
	\end{tabular}
\end{table}

\begin{figure}[!t]
	\centering
	\includegraphics[width=3.5in]{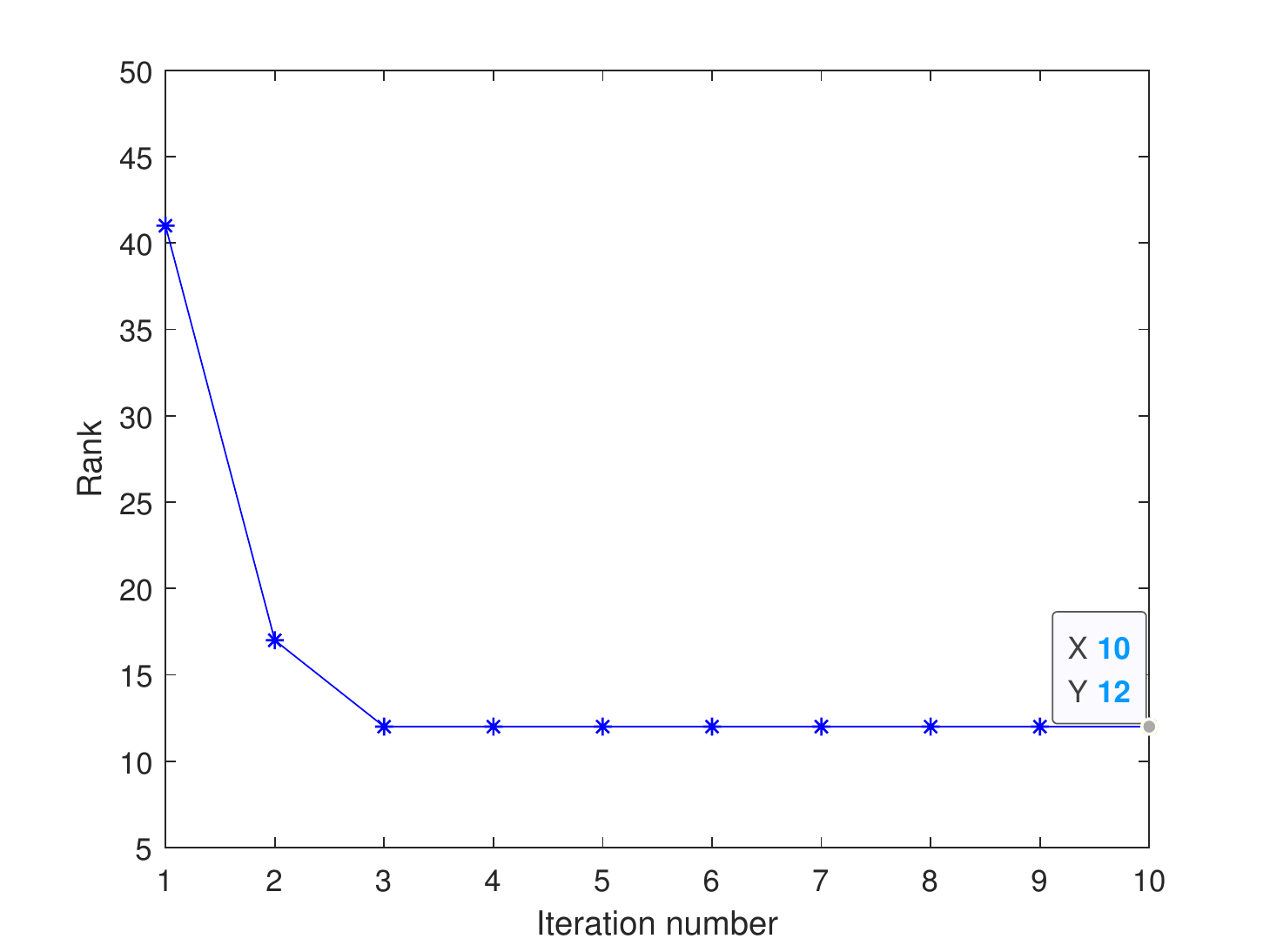}
	\caption{The change of rank with iteration times.}
	\label{fig3}
\end{figure}

In order to more clearly observe the change of rank as the number of iterations increases, the change curve is shown in Fig.\ref{fig3}.
In these ten iterations, we can clearly see that the rank of the hankel matrix (which also equals to the element number) decreases with the number of iterations, and then tends to be stable at 12.

\subsection{Chebshev weighting with notched sidelobes}
\begin{figure}[!tpb]
	\centering
	\subfigure[]
	{\includegraphics[width=3.5in]{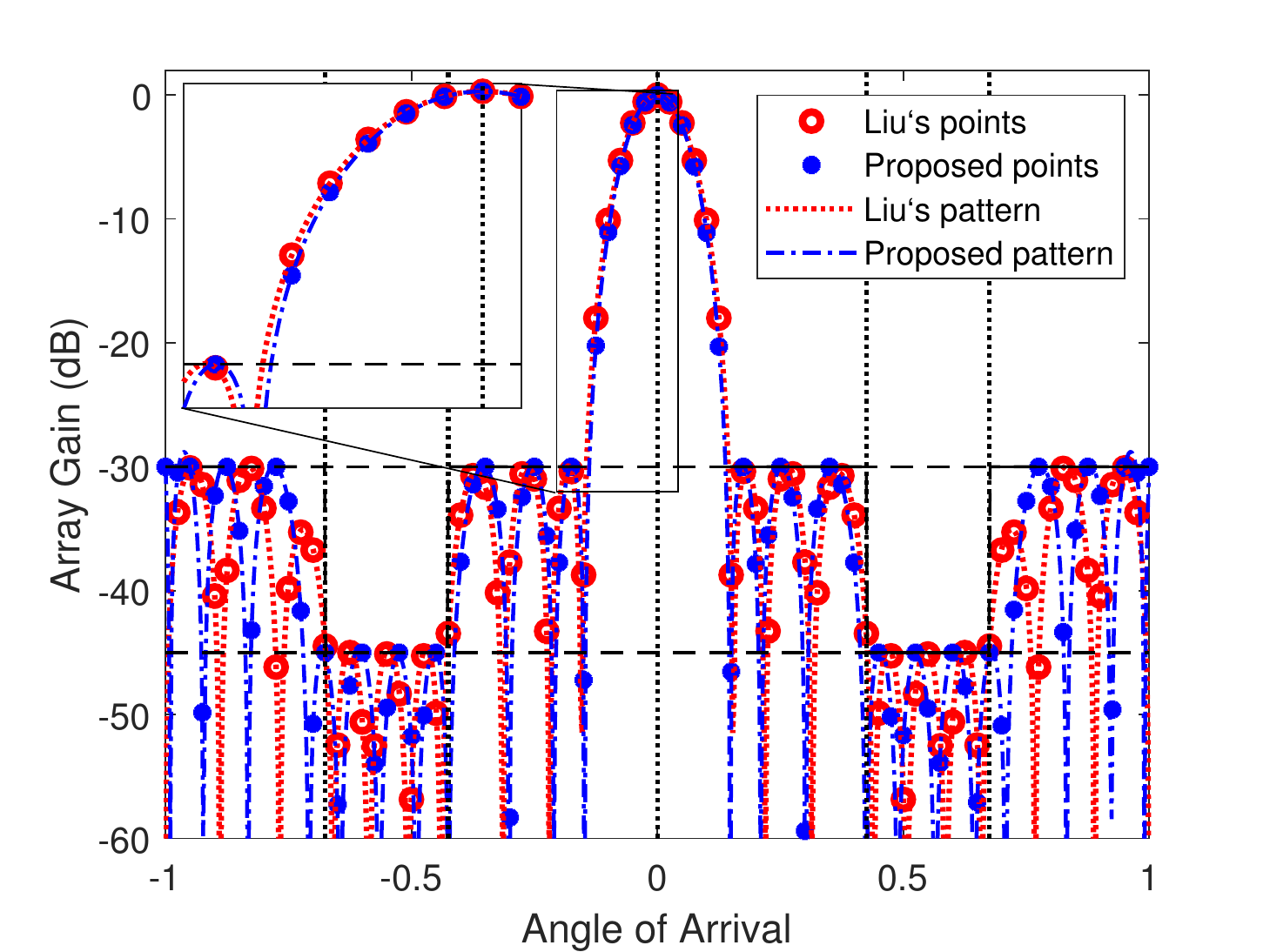}%
		\label{beamcom2}}
	\ \
	\subfigure[]
	{\includegraphics[width=3.5in]{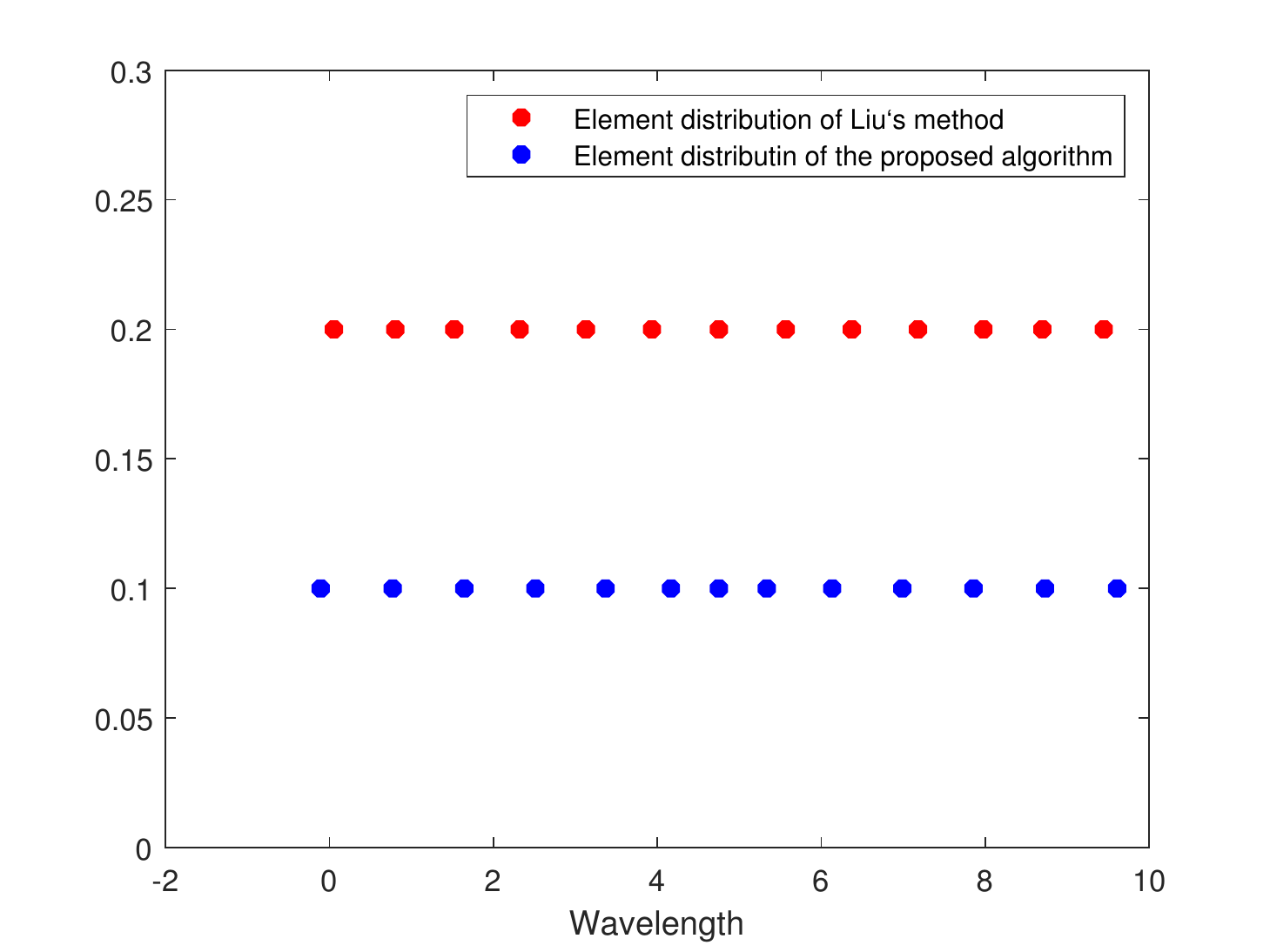}%
		\label{elecom2}}			
	\caption{Comparision between Liu's method and the proposed method under different sidelobe constraint.
		(a) Beam pattrens.
		(b) Elements distribution (both are 13 elements).}
	\label{fig4}
\end{figure}

After showing the basic effect of the proposed algorithm. We will give the simulation in a different scene in this subsection. We constrained the sidelobe to -30dB in the pre-subsection successfully, on this basis, apply another side lobe constraint of -45db within a specific angle range, it is worth mentioning that the notch on the side lobe needs to be set symmetrically to ensure the effectiveness of the algorithm.  The result is given in Fig.\ref{fig4}.

Similarly, we showed both the comparison of the beam pattern and element distribution. It can be seen that, even if the notch of the proposed algorithm is slightly wider than the preset notch width, we successfully applied two different constraints to the sidelobe region with the mainlobe maintaining its direction. In this situation, both methods reduce the number of array elements from 20 to 13. In the same way, because our method provides more degrees of freedom, it also shows some advantages in the generated pattern. Compared with MPM, the beam formed by our method has narrower main lobe. Like in subsection A, we also list the positions of the 13 new elements generated by the two methods (TABLE \ref{table2}) to see the difference of array element position directly.

\begin{table}[!t]
	\renewcommand{\arraystretch}{1.5}
	\caption{The table of element position and weight}
	\label{table2}
	\centering
	\begin{tabular}{c | c | c || c | c | c}
		\hline
		$ n $ & $ \hat{d_n} $ & $ d_n $  
		&$ n $ & $ \hat{d_n} $ & $ d_n $\\
		\hline	
		1 & -0.1074 &0.0546 &8 & 5.3345  &5.5664 \\
		2 & 0.7714  &0.8035 &9 & 6.1321  &6.3717 \\
		3 & 1.6444	&1.5227 &10& 6.9879  &7.1795 \\
		4 & 2.5121  &2.3205 &11& 7.8556  &7.9773 \\
		5 & 3.3679  &3.1283 &12& 8.7286  &8.6965 \\
		6 & 4.1655  &3.9336 &13& 9.0674  &9.4454 \\
		7 & 4.7500  &4.7500 \\
		\hline
	\end{tabular}
\end{table}

\subsection{Under different initial parameters}
\begin{figure}[!tpb]
	\centering
	\subfigure[]
	{\includegraphics[width=3.5in]{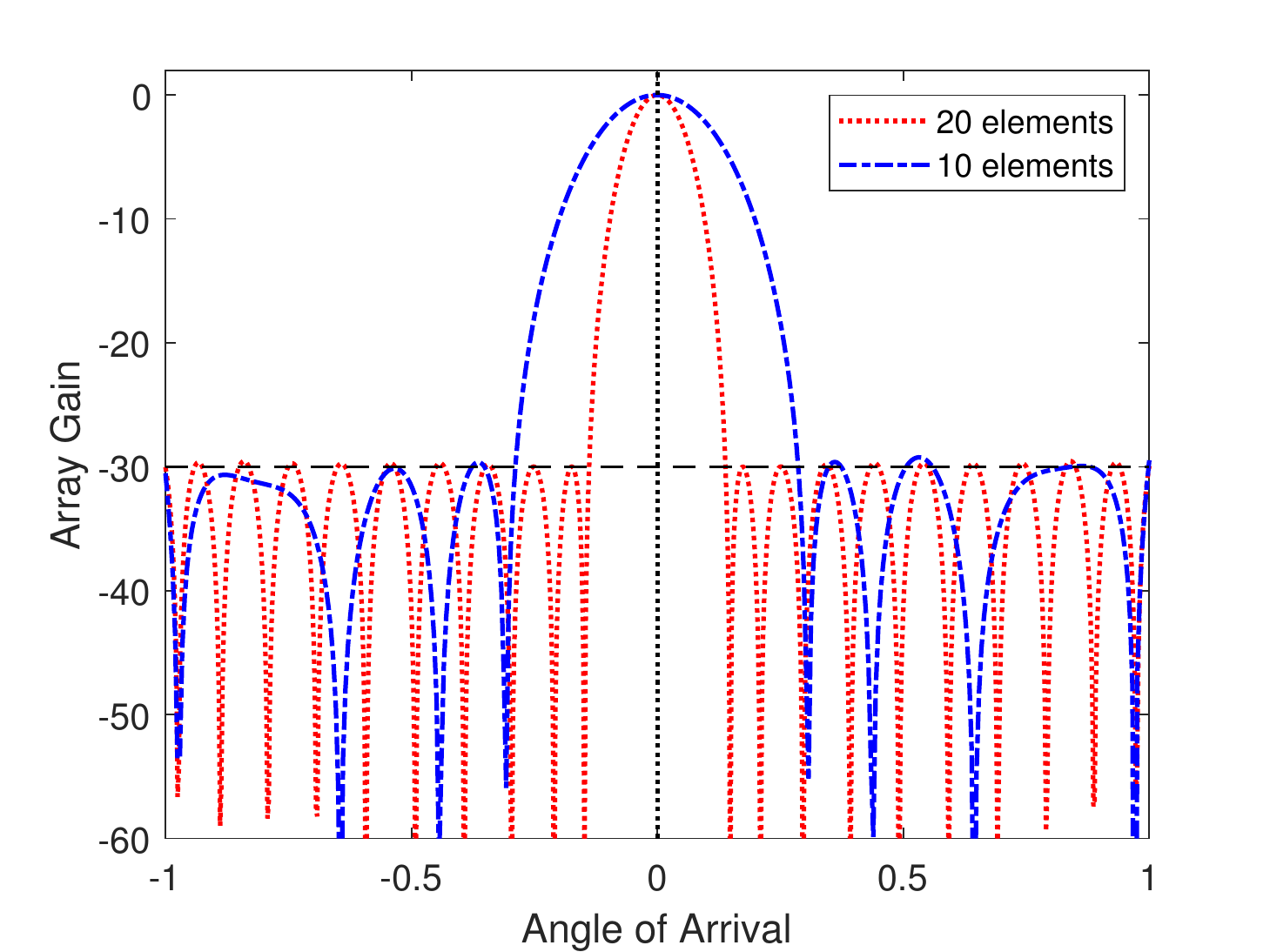}%
		\label{comele}}
	\ \
	\subfigure[]
	{\includegraphics[width=3.5in]{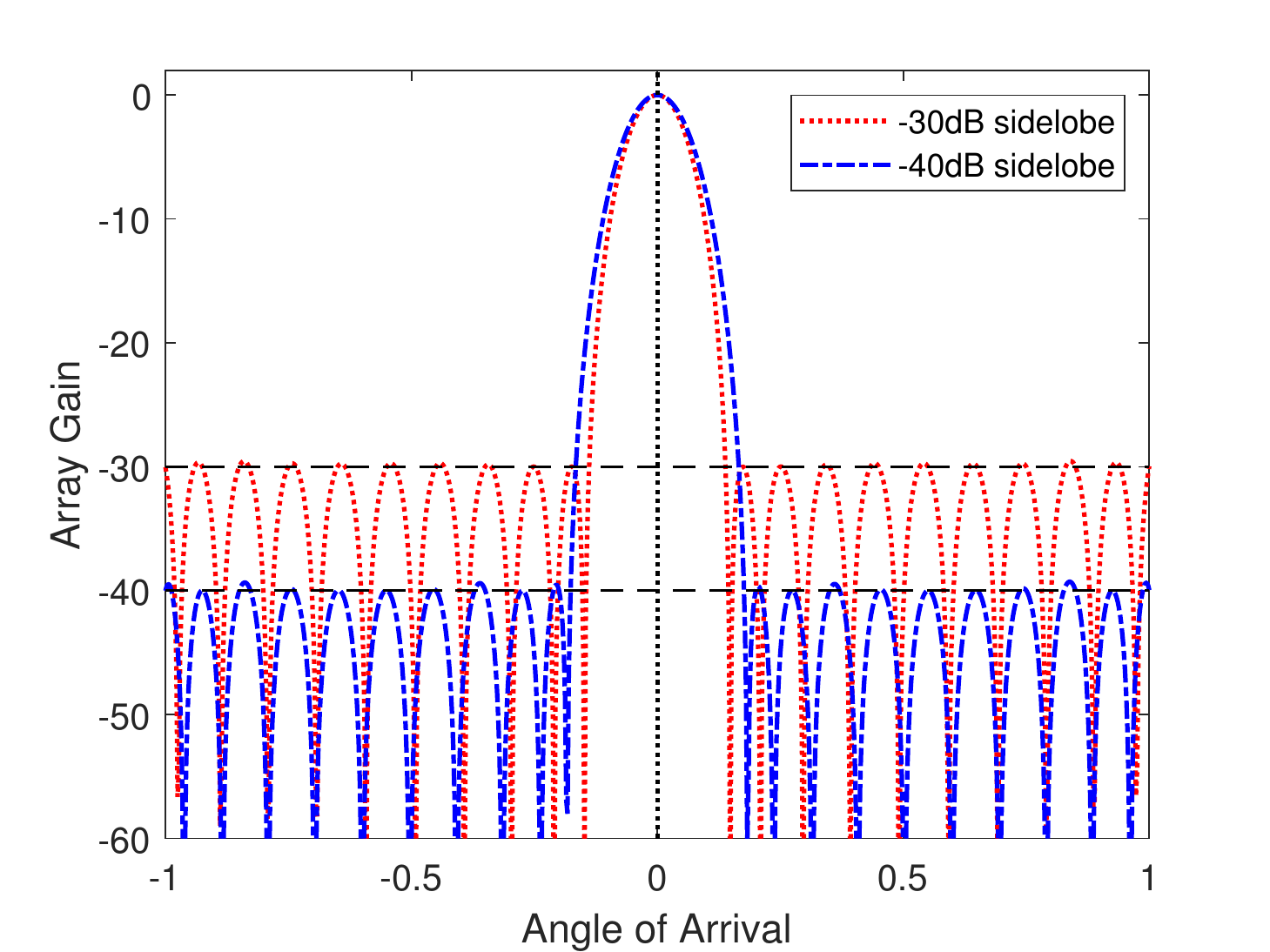}%
		\label{comside}}			
	\caption{Comparision between Liu's method and the proposed method.
		(a) Patterns with different element numbers.
		(b) Patterns with different sidelobe constraints.}
	\label{fig5}
\end{figure}

In this section, we will change the number of initial elements and the sidelobe constraint. In the same way, in addition to the middle sampling point, we also approach a sampling point on the left and right side of the mainlobe. Then the result is given in Fig.\ref{fig5}.
We can notice that the main lobe width becomes narrower with the increase of the initial number of elements in Fig.\ref{comele}.

Besides, we also change the level of the sidelobe constraint. The result is given in Fig.\ref{comside}.
We can notice that the main lobe width becomes wider as the side lobe constraint decreases.

\section{Conclusion}
In this paper, we have proposed a new algorithm to reduce the number of array elements, which has been simulated and compared with MPM algorithm. The simulation results in two different environments with side lobe constraints show that the ideal pattern has been obtained, and the non-uniform array has been completed, which has reduced the number of array elements. In contrast, when the number of array elements is the same, our method has some advantages under two different constraints, the side lobe constraint is better or the main lobe is narrower. However, this algorithm still has many shortcomings, such as the number of samples can not be too small, the beam pattern is best symmetrical. The problems like this need to be studied in depth to find algorithms with better performance and more adaptability.

%\balance
\bibliography{lowranksynthesisARXIV}
%\bibliography{refs}

\end{document}